\documentclass[preprint,aps,superscriptaddress,nofootinbib,tightenlines]{revtex4}

\usepackage{amsmath,amssymb}
\usepackage{graphicx}
\usepackage{bm}
\usepackage{comment}

\newcommand{\beq}{\begin{equation}}
\newcommand{\eeq}{\end{equation}}
\newcommand{\bea}{\begin{eqnarray}}
\newcommand{\eea}{\end{eqnarray}}

\def\OMIT#1{{}}

\newcommand{\lsim}{\ \raisebox{-0.7ex}{$\stackrel{\textstyle <}{\sim}$ }}

\newcommand{\Choose}[2]{{^{#1}C_{#2}}}

\def\abar{\overline{a}}

\def\sheep{$\overline{\overline{\eta}}_{3K^-}^L$}
\def\sheeppi{$\overline{\overline{\eta}}_{3\pi^-}^L$}
\def\apipi{\overline{a}_{\pi\pi}^{(I=2)}}
\def\aKK{\overline{a}_{KK}^{(I=1)}}

\begin{document}
\begin{figure}[!t]
  \vskip -1.6cm
  \leftline{\includegraphics[width=0.25\textwidth]{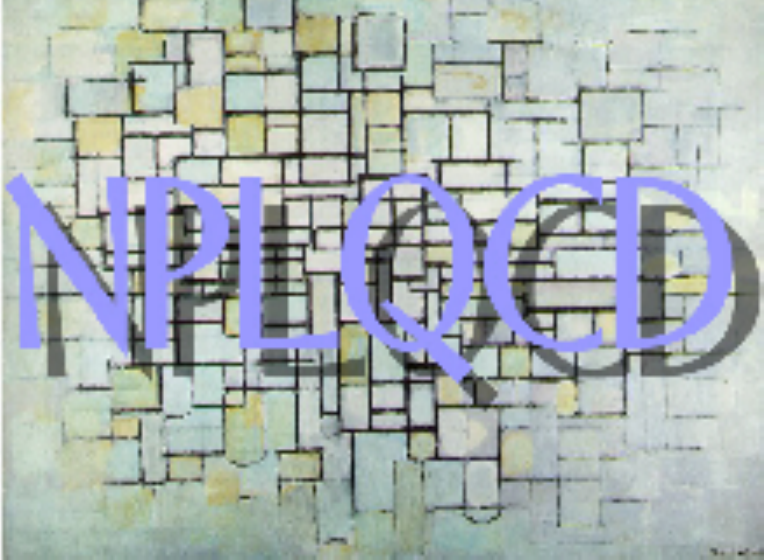}}
\end{figure}

\preprint{\vbox{ \hbox{JLAB-THY-08-849}
    \hbox{NT@UW-08-13}   
  \hbox{UMD-40762-414}
  }}

\vskip .5cm

\title{Kaon Condensation with Lattice QCD}

\vskip .5cm 
\author{William Detmold}
\affiliation{Department of Physics, University of Washington, 
Seattle, WA 98195-1560.}
\author{Kostas Orginos}
\affiliation{Department of Physics, College of William and Mary, Williamsburg,
  VA 23187-8795.}
\affiliation{Jefferson Laboratory, 12000 Jefferson Avenue, 
Newport News, VA 23606.}
\author{Martin J.~Savage}
\affiliation{Department of Physics, University of Washington, 
Seattle, WA 98195-1560.}
\author{Andr\'e Walker-Loud}
\affiliation{Department of Physics, College of William and Mary, Williamsburg,
  VA 23187-8795.}
\affiliation{Department of Physics, University of Maryland, College Park, MD
  20742-4111.}
\collaboration{ NPLQCD Collaboration }
\noaffiliation
\vphantom{}

\date{\today}

\vskip 0.8cm
\begin{abstract}
\noindent 
Kaon condensation may play an important role in the structure of
hadronic matter at densities greater than that of nuclear matter, as
exist in the interior of neutron stars.  We present the results of the
first lattice QCD investigation of kaon condensation obtained by
studying systems containing up to twelve negatively charged kaons.
Surprisingly, the properties of the condensate that we calculate are
remarkably well reproduced by leading order chiral perturbation
theory.  In our analysis, we also determine the three-kaon interaction
from the multi-kaon systems and update our results for pion
condensates.

\end{abstract}
 \maketitle

\vfill\eject

%%%%%%%%%%%%%%%%%%%%%%%%%%%%%%%%%%%%%%%%%%%%%%%%%%%
%
%               Intro
%
%%%%%%%%%%%%%%%%%%%%%%%%%%%%%%%%%%%%%%%%%%%%%%%%%%%
\section{Introduction 
\label{sec:Intro}}

\noindent 
A key ingredient in determining whether supernovae evolve into
black-holes or neutron stars is the nuclear equation of state
(NEOS)~\cite{Brown:1993jz}.  The NEOS depends upon the degrees of
freedom at a given density, which in turn are determined by the masses
of the various hadrons in the medium and their interactions. One
possibility is that it is more energetically favorable for a
Bose-Einstein condensate of negatively charged mesons to form in the
hadronic matter.  Naively, one expects pion condensation to occur
before kaon condensation\footnote{ A second possibility is that it is
  energetically favorable to have $\Sigma^-$'s present in the nuclear
  material in addition to neutrons and protons. The in-medium mass of
  the $\Sigma^-$ is expected to be significantly less than in
  free-space due to its interactions with neutrons, but the precise
  mass-shift is uncertain due to the model dependence of existing
  theoretical calculations~\cite{Page:2006ud}. Experimentally, little
  is known about the $\Sigma^-$-neutron interaction and a precise
  Lattice QCD calculation of these interactions~\cite{Beane:2006gf}
  would greatly reduce the uncertainty in this particular contribution
  to the NEOS.}, however, the repulsive pion-nucleon interactions and
the attractive kaon-nucleon interaction lead to kaon condensation
being the prime candidate for softening the NEOS at moderate nuclear
densities.  In theoretical explorations of kaon condensation, chiral
perturbation theory ($\chi$PT) is used to describe the kaon-nucleon
interactions and the kaon self-interactions at leading order (LO) or
next-to-leading order (NLO) in the chiral
expansion~\cite{KaplanNelson}.  The energy of the hadronic plus
electronic system is minimized with respect to the vacuum expectation
values (the condensates) of the meson fields subject to the
constraints imposed by charge neutrality and the conservation of
baryon number.  In the case of non-interacting mesons, once the
chemical potential exceeds the rest mass of a given species of meson a
condensate will form with an infinite density of mesons.  However, in
the case of a meson with repulsive self-interactions, the number
density of mesons will remain finite and a smooth function of the
chemical potential.  Depending upon the many-body technique employed
to describe the system, a condensate of $K^-$'s may become
energetically favorable at baryon densities, $n_b$, as low as three
times nuclear matter density (for a nice discussion, see
ref.~\cite{Page:2006ud}).  

In this work we perform the first Lattice QCD studies of kaon
condensation.  In particular, we calculate the correlation functions
of systems containing up to twelve ($n=12$) $K^-$'s on the coarse MILC
lattices with spatial extent $L\sim 2.5~{\rm fm}$ and $L\sim 3.5~{\rm
  fm}$, both with lattice spacing $b\sim 0.125~{\rm fm}$, and on the
fine MILC lattices with spatial extent $L\sim 2.5~{\rm fm}$ and
lattice spacing $b\sim 0.090~{\rm fm}$.  LO $\chi$PT is found to
describe the chemical potential, and its density dependence,
remarkably well over the densities explored.  This work is an
extension of our previous work on charged-pion
condensates~\cite{Beane:2007es,Detmold:2008fn} and, for completeness,
we update that analysis with the new ensembles studied herein.

%%%%%%%%%%%%%%%%%%%%%%%%%%%%%%%%%%%
\section{Condensation of 
  Pseudo-Goldstone Bosons in $\chi$PT}
\label{sec:XPT}

At LO in $\chi$PT, the Lagrange density describing the low-energy
dynamics of the pseudo-Goldstone bosons associated with the
spontaneous breaking of the approximate $SU(3)_L\otimes SU(3)_R$
chiral symmetry of QCD has the form
\begin{eqnarray}
{\cal L} & = & 
{f^2\over 8}\ {\rm  Tr}\left[\ \partial_\mu\Sigma\ \partial^\mu\Sigma^\dagger\ \right]
\ +\ \lambda\ {\rm  Tr}\left[\ m_q\ \left(\Sigma\ +\ \Sigma^\dagger\ \right)\
\right]
\ \ \ ,
\label{eq:lochipt}
\end{eqnarray}
where 
$\Sigma$ is the exponential of the meson fields,
\begin{eqnarray}
\Sigma & = & e^{i 2 M/ f}
\ \ ,\ \ 
M\ =\ 
\left(
\begin{array}{ccc}
{1\over\sqrt{2}}\pi^0 + {1\over\sqrt{6}}\eta^0 & \pi^+ & K^+\\
\pi^- & -{1\over\sqrt{2}}\pi^0 + {1\over\sqrt{6}}\eta^0 &K^0\\
K^- & \overline{K}^0 & -{2\over\sqrt{6}}\eta^0 
\end{array}
\right)
\ \ \ ,
\end{eqnarray}
and where $f$ is the chiral limit of the pion decay constant (in this
normalization convention the physical pion decay constant is
$f_\pi\sim 132~{\rm MeV}$).  At LO in the chiral expansion, the quark
mass matrix, $m_q$, can be written in terms of the meson masses,
\begin{eqnarray}
\lambda\ m_q & = & {f^2\over 8}
\left(
\begin{array}{ccc}
m_\pi^2 & 0 &  0\\
0 & m_\pi^2 & 0 \\
0&0& 2 m_K^2-m_\pi^2
\end{array}
\right)
\ \ \ .
\end{eqnarray}
The Lagrange density in eq.~(\ref{eq:lochipt}), in addition to
describing the free-field dynamics of the mesons, also contains
interactions containing arbitrary numbers of mesons.  At zero
momentum, the LO contributions to the interactions between three
charged pions or between three charged kaons are given by
\begin{eqnarray}
{\cal L} & = & 
{2\over 3 f^4} 
\left[\ 
 m_\pi^2 (\pi^+)^3(\pi^-)^3\ +\ 
m_K^2 (K^+)^3 (K^-)^3\ 
\right]
\ \ \ .
\end{eqnarray}
This contribution provides the naive dimensional analysis (NDA)
estimate of the size of the three-meson
interaction~\cite{Beane:2007qr,Detmold:2008gh} of $1/(m_\pi f^4)$ for
the $\pi^-\pi^-\pi^-$ interaction and $1/(m_K f^4)$ for the
$K^-K^-K^-$ interaction in the non-relativistic theory describing the
very-low momentum interactions between the mesons~\footnote{The
  relativistic and non-relativistic fields differ in their
  normalization by a factor of $\sqrt{m}$.}.

In the presence of a chemical potential for the third component of
isospin, $\mu_I$, and strangeness, $\mu_S$, the partial derivatives in
eq.~(\ref{eq:lochipt}) become covariant derivatives
\cite{Son:2000xc,Son:2000by,Kogut:2001id}
\begin{eqnarray}
\partial_\mu\Sigma \rightarrow D_\mu\Sigma = \partial_\mu\Sigma + i \delta_{\mu0} \mu_I
\left[\hat I_z , \Sigma\right]\ +\  i \delta_{\mu0} \mu_S \left[\hat S , \Sigma\right]
\ \ \ ,
\end{eqnarray}
where
\begin{eqnarray}
\hat I_z & = & 
{1\over 2}
\left(
\begin{array}{ccc}
+1&0&0\\0&-1&0\\0&0&0
\end{array}
\right)
\ \ ,\ \ 
 \hat S\ =\ 
\left(
\begin{array}{ccc}
0&0&0\\0&0&0\\0&0&+1
\end{array}
\right)
\ \ \ .
\end{eqnarray}
In the situation where $\mu_{\pi^-}=-\mu_I > m_{\pi}$ and
$\mu_{K^-}=\mu_S-{1\over 2}\mu_I < m_{K}$, it is energetically
favorable to form a $\pi^-$ condensate but not a $K^-$ condensate, and
at LO in $\chi$PT the number density of $\pi^-$'s
is~\cite{Son:2000xc,Son:2000by}
\begin{eqnarray}
\rho_{\pi^-} & = & 
{f^2_{\pi^-}\over 2} \ \left(\ \mu_{\pi^-}\ -\ {m_\pi^4\over \mu_{\pi^-}^3}
\ \right)
\ \ \ .
\label{eq:pionCOND}
\end{eqnarray}
The analysis leading to this result easily extended to
kaon-condensates \cite{Kogut:2001id}.  For $\mu_{\pi^-} < m_{\pi}$ and
$\mu_{K^-} > m_{K}$ there will be a condensate of $K^-$'s but not of
$\pi^-$'s~\footnote{Mixed states with different numbers of pions and
  kaons are the subject of ongoing work.}.  For chemical potentials in
this range, the number density of $K^-$'s as a function of the
chemical potential, $\mu_{K^-}$, is, at LO in $\chi$PT,
\begin{eqnarray}
\rho_{K^-} & = &
{f^2_{K^-}\over 2} \ \left(\ \mu_{K^-}\ -\ {m_K^4\over \mu_{K^-}^3}
\ \right)
\ \ \ .
\label{eq:kaonCOND}
\end{eqnarray}
In eqs.~(\ref{eq:pionCOND}) and (\ref{eq:kaonCOND}), we have used LO
$\chi$PT relations to express the result in terms of the physical
masses and decay constants.  It is this relation that we will compare
with Lattice QCD calculations.  The reason for ``testing'' the
relation in eq.~(\ref{eq:kaonCOND}) is that, given the relatively
large expansion parameters of $SU(3)_L\otimes SU(3)_R$ $\chi$PT, it is
natural to expect that there will be corrections to $\rho_{K^-}$, and
hence to the dependence of $\mu_{K^-}$ on $\rho_{K^-}$, at the level
of $m_K^2/\Lambda_\chi^2 \sim 25\%$, where $\Lambda_\chi\sim1$~GeV is
the scale of chiral symmetry breaking.

In order to assess the relevance of the present work, it is
instructive to compare the value of the $K^-$-condensates we calculate
to those found to be relevant to the dense nuclear matter in the
interior of neutron stars.  Decomposing the charged kaon fields into
their real and imaginary parts, $K^\pm = \left(K_1 \mp i
  K_2\right)/\sqrt{2}$, and denoting the condensates $\langle
K_1\rangle = v_1$ and $\langle K_2\rangle = v_2$ and further defining
$\langle K\rangle = \sqrt{v_1^2+v_2^2}/\sqrt{2}$, the condensate is
related to the chemical potential via
\begin{eqnarray}
{ \langle K\rangle \over f_K} & = & {1\over 2}\cos^{-1}\left({m_K^2\over
    \mu_{K^-}^2}\right)
\ \ \ .
\label{eq:kaonCONDvev}
\end{eqnarray}
Condensates of magnitude $\langle K\rangle/f_K \lsim 0.3$ are found
to be possible at the densities in the interior of neutron
stars~\cite{KaplanNelson}.  This corresponds to $\mu_{K^-}/m_K -1
\lsim 0.1$, which is contained in the region explored in the Lattice
QCD calculations described in this work.

%%%%%%%%%%%%%%%%%%%%%%%%%%%%%%%%%%%%%%%%%%%%%%%%%%%%%%%
\section{Multi-Meson Energies in a Finite Volume}
\label{sec:multi-meson-energies}

\noindent
In recent works \cite{Beane:2007qr,Detmold:2008gh,Tan:2007bg}, the
analytic volume dependence of the ground-state energy of $n$ identical
bosons in a periodic volume of side $L$ has been computed to ${\cal
  O}(L^{-7})$, extending the classic results of Bogoliubov \cite{Bog}
and Lee, Huang and Yang \cite{Lee:1957}.  The resulting shift in
ground-state energy of $n$ particles of mass $M$ due to their
interactions is \cite{Detmold:2008gh}
\begin{eqnarray}
 \Delta E_n &=&
  \frac{4\pi\, \abar}{M\,L^3}\Choose{n}{2}\Bigg\{1
-\left(\frac{\abar}{\pi\,L}\right){\cal I}
+\left(\frac{\abar}{\pi\,L}\right)^2\left[{\cal I}^2+(2n-5){\cal J}\right]
\nonumber 
\\&&\hspace*{2cm}
-
\left(\frac{\abar}{\pi\,L}\right)^3\Big[{\cal I}^3 + (2 n-7)
  {\cal I}{\cal J} + \left(5 n^2-41 n+63\right){\cal K}\Big]
\nonumber
\\&&\hspace*{2cm}
+
\left(\frac{\abar}{\pi\,L}\right)^4\Big[
{\cal I}^4 - 6 {\cal I}^2 {\cal J} + (4 + n - n^2){\cal J}^2 
+ 4 (27-15 n + n^2) {\cal I} \ {\cal K}
\nonumber\\
&&\hspace*{4cm}
+(14 n^3-227 n^2+919 n-1043) {\cal L}\ 
\Big]
\Bigg\}
\nonumber\\
&&
+\ \Choose{n}{3}\left[\ 
{192 \ \abar^5\over M\pi^3 L^7} \left( {\cal T}_0\ +\ {\cal T}_1\ n \right)
\ +\ 
{6\pi \abar^3\over M^3 L^7}\ (n+3)\ {\cal I}\ 
\right]
\nonumber\\
&&
+\ \Choose{n}{3} \ {1\over L^6}\ \overline{\overline{\eta}}_3^L\ 
\ \ + \ {\cal O}\left(L^{-8}\right)
\ \ \ \ ,
\label{eq:energyshift}
\end{eqnarray}
where the parameter $\abar$ is related to the scattering
length\footnote{In this work we use the Nuclear Physics sign
  convention for the scattering length, opposite to that used in
  Particle Physics.  In this convention, the $\pi^+\pi^+$ scattering
  length is positive, corresponding to a repulsive interaction.}, $a$,
and the effective range, $r$, by
\begin{eqnarray}
a
& = & 
\overline{a}\ -\ {2\pi\over L^3} \overline{a}^3 r \left(\ 1 \ -\
  \left( {\overline{a}\over\pi L}\right)\ {\cal I} \right)\ \ .
\label{eq:aabar}
\end{eqnarray}
The geometric constants that enter into eq.~(\ref{eq:energyshift}) are
\begin{align} 
  &{\cal I}\ =\ -8.9136329\,, &{\cal J}\ =\ 16.532316\,, \qquad\qquad
  {\cal K}\ = \ 8.4019240\,,
  \nonumber\\
  &{\cal L}\ = \ 6.9458079\,, &{\cal T}_0\ = -4116.2338\,,
  \qquad\qquad {\cal T}_1\ = \ 450.6392\,, &
\label{eq:sums}
\end{align}
and $^nC_m=n!/m!/(n-m)!$.  The three-body contribution to the
energy-shift given in eq.~(\ref{eq:energyshift}) is represented by the
parameter $\overline{\overline{\eta}}_3^L$, which is a combination of
the volume-dependent, renormalization group invariant quantity,
$\overline{\eta}_3^L$, and contributions from the two-body scattering
length and effective range,
\begin{eqnarray}
\overline{\overline{\eta}}_3^L & = & \overline{\eta}_3^L  \left(\ 1 \ -\
  6  \left({\overline{a}\over\pi L}\right)\ {\cal I} \right)
\ +\ {72\pi \overline{a}^4 r\over M L} \ {\cal I}
\ \ \ ,
\label{eq:eta3barbar}
\end{eqnarray}
where
\begin{align} 
  \overline{\eta}_3^L & = \eta_3(\mu)\ +\ {64\pi a^4\over
    M}\left(3\sqrt{3}-4\pi\right)\ \log\left(\mu L\right)\ -\ {96
    a^4\over\pi^2 M} {\cal S}_{\rm MS} \ \ \ .
\label{eq:etathreebar}
\end{align}
The quantity $\eta_3(\mu)$ is the 
renormalization scale dependent
coefficient of the three-$\pi^+$
interaction that appears in the effective Hamiltonian density
describing the system \cite{Detmold:2008gh}. The quantity ${\cal S}$ is renormalization
scheme dependent and its value in the minimal subtraction (MS)
scheme is ${\cal S}_{\rm MS}\ = \ -185.12506$.

For $n=2$, the last two terms in eq.~(\ref{eq:energyshift}) vanish and
the remaining terms constitute the small $\abar/L$ expansion of the
exact eigenvalue equation derived by L\"uscher
\cite{Luscher:1986pf,Luscher:1990ux},
\begin{eqnarray}
p\cot\delta(p) \ =\ {1\over \pi L}\ {\bf
  S}\left(\,\left(\frac{p L}{2\pi}\right)^2\,\right)
\ \ ,
\label{eq:energies}
\end{eqnarray}
which is valid below the inelastic threshold, where 
$p\cot\delta(p)$ is the real part of the inverse scattering amplitude.
The regulated three-dimensional sum is~\cite{Luscher:1986pf,Luscher:1990ux,Beane:2003da}
\begin{eqnarray}
{\bf S}\left(\, x \, \right)\ \equiv \ \sum_{\bf j}^{ |{\bf j}|<\Lambda}
{1\over |{\bf j}|^2-x}\ -\  {4 \pi \Lambda}
\ \ \  ,
\label{eq:Sdefined}
\end{eqnarray}
where the summation is over all triplets of integers ${\bf j}$ such
that $|{\bf j}| < \Lambda$ and the limit $\Lambda\rightarrow\infty$ is
implicit.

%%%%%%%%%%%%%%%%%%%%%%%%%%%%%%%%%%%%%%%%%%%%%%%%%%%
%
%               Lattice Details
%
%%%%%%%%%%%%%%%%%%%%%%%%%%%%%%%%%%%%%%%%%%%%%%%%%%%
\section{Methodology and Details of the Lattice Calculation \label{sec:Method}}

\noindent
The computation in this paper uses the mixed-action lattice QCD scheme
developed by LHPC~\cite{Renner:2004ck,Edwards:2005kw} and described
fully in Ref.~\cite{Detmold:2008fn} where multi-pion systems are
investigated in detail.  The present calculations were performed
predominantly with the coarse,
asqtad-improved~\cite{Orginos:1999cr,Orginos:1998ue} MILC
configurations generated with rooted staggered sea
quarks~\cite{Bernard:2001av}\footnote{The results of this paper assume
  that the fourth-root trick recovers the correct continuum limit of
  QCD.} with a lattice spacing of $b\sim 0.125$~fm, and a spatial
extent of $L\sim 2.5$~fm.  These configurations were
HYP-smeared~\cite{Hasenfratz:2001hp,DeGrand:2002vu,DeGrand:2003in,Durr:2004as}
and used to generate domain-wall fermion
propagators~\cite{Kaplan:1992bt,Shamir:1992im,Shamir:1993zy,Shamir:1998ww,Furman:1994ky}.
The strange quark was held fixed near its physical value while the
degenerate light quarks were varied over a range of
masses\footnote{The mass of the valence quark was chosen to yield a
  pion mass that matches the pseudo-Goldstone staggered pion
  \cite{Detmold:2008fn}.}; see Table~\ref{tab:MILCcnfs} and
Refs.~\cite{Beane:2006mx,Beane:2006pt,Beane:2006fk,Beane:2006kx,Beane:2006gf}
for details.  Further calculations were performed using one fine MILC
ensemble, with $b\sim 0.090~{\rm fm}$ and $L\sim 2.5$~fm and on a
larger volume coarse MILC ensemble with $b\sim 0.125$~fm and $L\sim
3.5$~fm.  On the coarse, 2.5~fm MILC lattices, Dirichlet boundary
conditions were implemented to reduce the original time extent of 64
time-slices down to 32 and thus save in computational time.  While
this procedure leads to minimal degradation of baryon signals, it does
limit the number of time slices available for fitting meson
properties. For the fine MILC ensemble and the large volume coarse
MILC ensemble, anti-periodic boundary conditions in time were
implemented and all time slices are available for analysis.
%%%%%%%%%%%%%%%%%%%%%%%%%%%%%%%%%%%%%%%%%%%%%%%%%%%
%
%               TABLE: Lattice Parameteres
%
%%%%%%%%%%%%%%%%%%%%%%%%%%%%%%%%%%%%%%%%%%%%%%%%%%%
\begin{table}[t]
 \caption{The parameters of the MILC gauge configurations and
   domain-wall quark propagators used in this work. The subscript $l$
   denotes light quark (up and down), and  $s$ denotes the strange
   quark. The superscript $dwf$ denotes the bare-quark mass for the
   domain-wall fermion propagator calculation. The last column is the 
   number of configurations times the number of sources per
   configuration. For the ensembles labeled with ``P$\pm$A'', propagators
   that were periodic in the temporal direction were computed in
   addition to those with anti-periodic temporal boundary conditions
   ($b m_{res}$ was not computed for these sets but is expected to be
   very close to the result from the corresponding set of propagators
   with anti-periodic temporal boubdary conditions).}
\label{tab:MILCcnfs}
\begin{ruledtabular}
\begin{tabular}{ccccccc}
 Ensemble        
&  $b m_l$ &  $b m_s$ & $b m^{dwf}_l$ & $ b m^{dwf}_s $ & $10^3 \times b
m_{res}$ & \# of propagators   \\
\hline 
2064f21b676m007m050 &  0.007 & 0.050 & 0.0081 & 0.081  & 1.604 & 1038\ $\times$\ 24 \\
2064f21b676m010m050 &  0.010 & 0.050 & 0.0138 & 0.081  & 1.552 & 768\ $\times$\ 24 \\
2064f21b679m020m050 &  0.020 & 0.050 & 0.0313 & 0.081  & 1.239 & 486\ $\times$\ 24 \\
2064f21b681m030m050 &  0.030 & 0.050 & 0.0478 & 0.081  & 0.982 & 564\ $\times$\ 20 \\
\hline
2896f2b709m0062m031 & 0.0062 & 0.031 & 0.0080 & 0.0423 & 
\ 0.380 \  & 1001\ $\times$\ 7 \\
2896f2b709m0062m031 P$\pm$A& 0.0062 & 0.031 & 0.0080 & 0.0423 & 
\ --- \  & 1001\ $\times$\ (1+1) \\
\hline
2864f2b676m010m050 & 0.010 & 0.050 & 0.0138 & 0.081 & 
\ 1.552 \  & 137 \ $\times$\ 8 \\
2864f2b676m010m050 P$\pm$A & 0.010 & 0.050 & 0.0138 & 0.081 & 
\ --- \  & 274 \ $\times$\ (2+2) \\
\end{tabular}
\end{ruledtabular}
\end{table}
%%%%%%%%%%%%%%%%%%%%%%%%%%%%%%%%%%%%%%%%%%%%%%%%%%%
%
In order to extract precise measurements from the configurations,
propagators were computed from multiple sources displaced both
temporally and spatially on each configuration  The correlators were blocked
so that one averaged correlator per configuration was used in the
subsequent statistical analysis.  Further blocking over sets of
configurations did not change the resulting uncertainties associated
with each observable.

For the fine MILC ensemble and the large volume coarse MILC ensemble,
additional propagators were computed using periodic temporal boundary
conditions for a subset of the source points. Combining periodic and
anti-periodic (in time) propagators from the same source leads to a
propagator that is periodic in time over twice the temporal extent of
the lattice. These ensembles, labeled ``P$\pm$A'' in
Table~\ref{tab:MILCcnfs}, served to explore the effects of temporal
boundary conditions.\footnote{ For the correlators of systems
  containing multiple hadrons (which are extremely sensitive to
  truncation errors in arithmetic operations), we find that performing
  the P$\pm$A additions and subtractions in machine (double) precision
  leads to mild degradation of signals that worsens with increasing
  $n$. This indicates that differences in the iterative (conjugate
  gradient) propagator inversion for the two types of boundary
  conditions are becoming significant.  }

In order to determine the interaction energy in multi-meson systems,
both the single-meson, $C_{1}(t)$, and $n$-meson, $C_{n }(t)$,
correlation functions were computed, where $t$ is the Euclidean time
separation between the hadronic source and sink operators.  For
simplicity, meson correlation functions involving only zero-momentum
states were calculated.  The single-kaon correlation function is
\begin{equation}\label{eq:C_K}
        C_{K^-}(t) = \sum_\mathbf{x} \langle\, K^+(t,\mathbf{x})\
        K^-(0,\mathbf{0})\, \rangle\, , 
\end{equation}
where the sum over all spatial sites projects onto the zero-momentum
state, $\mathbf{p}=0$.  A correlation function which projects onto the
$n$ $K^-$ ground-state is
\begin{equation}\label{eq:C_nK}
        C_{n K^-}(t) = 
                 \langle\,  \left( \ \sum_{\bf x } K^+(t,{\bf x})\ \right)^n \
                 \left(\ \phantom{\sum_{\bf x }} \hspace{-14pt} K^-(0,
                   {\bf 0})\ \right)^n \, \rangle\, , 
\end{equation}
which can be constructed from one light and one strange quark propagator.
In eqs.~\eqref{eq:C_K} and \eqref{eq:C_nK}, $K^+(t,\mathbf{x}) =
\bar{s}(t,\mathbf{x}) \gamma_5 u(t,\mathbf{x})$ is a Gaussian-smeared
interpolating field for the charged kaon.  In the relatively-large
spatial volumes used in the calculation, the interaction energy is a
small fraction of the total energy that is dominated by the 
mass of the kaon.  
To determine this energy, the ratio of correlation functions,
\begin{equation}\label{eq:C_nK_C_K}
        G_{n K^-}(t) \equiv \frac{C_{n K^-}(t)}{ \left[\ C_{K^-}(t)\ \right]^n
        }\ \longrightarrow \  \mathcal{A}_n \ e^{-\Delta E_{nK} t}\, ,
\end{equation}
was constructed, with the arrow denoting the large-time,
infinite-number-of-gauge-configurations limit (far from the boundary).
Due to the (anti-)periodic boundary-conditions imposed on the
propagators computed on the fine, and large-volume lattices, the $K^-$
correlation function becomes a single $\cosh$ function far from the
source, while the $n$-kaon correlation functions become sums of
multiple $\cosh$'s, leading to a non-trivial form for $G_{n K^-}(t)$;
further details are given in Appendix \ref{sec:effects-temp-bound}. As
an alternative method to calculating the interaction energy (and a
check of the systematics), a Jackknife analysis of the difference
between the energies extracted from the long-time behavior of the
multiple- and single-kaon correlation functions individually was
performed, finding results in agreement with those determined from
eq.~\eqref{eq:C_nK_C_K}.  The interaction energy is related to the
$n$-particle energy eigenvalues, $E_{n K}$, and the kaon mass,
\begin{equation}\label{eq:DeltaE}
        \Delta E_{n K}\  \equiv \ E_{n K} - n\ m_K\, .
\end{equation}
A comprehensive review of the lattice techniques we have used in this
work can be found in Refs.~\cite{Detmold:2008fn,NPLQCDreview}.

\begin{figure}[!ht]
  \centering
  \includegraphics[width=0.98\columnwidth]{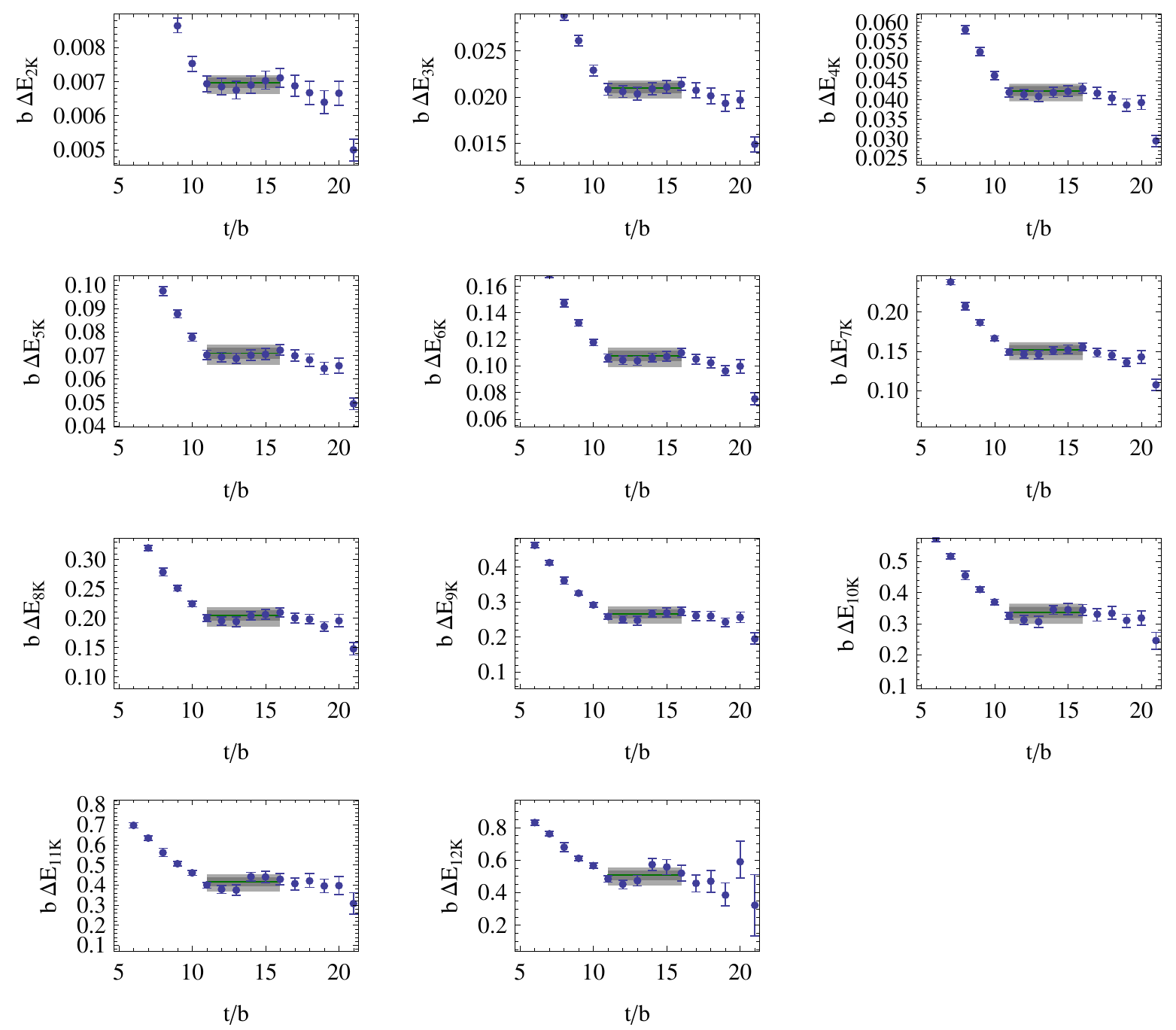}
  \caption{The energy-differences (in lattice units), $b\ \Delta E_{n K}$, for the
    multi-kaon systems with $m_\pi\sim 291~{\rm MeV}$.  The lighter
    (darker) shaded regions correspond to the statistical and
    systematic uncertainties combined in quadrature (statistical
    uncertainty) obtained from fitting the plateau regions.  The green
    line denotes the central value of the fit. }
  \label{fig:DeltaE007}
\end{figure}
To indicate the quality of the results of this calculation, in
fig.~\ref{fig:DeltaE007} we show the effective energy differences,
$\Delta E_{n K}$, as defined in eq.~(\ref{eq:DeltaE}), for the coarse
2.5 fm ensemble with $m_\pi\sim 291~{\rm MeV}$.  For each $n$ there is
clear, multi-time-slice, plateau from which to extract $\Delta E_{n
  K}$.  Figure ~\ref{fig:DeltaE007} shows that the systematic
uncertainties in the plateau region tend to become more significant as
the number of kaons in the system is increased, likely due to the fact
that the same propagators are being raised to high powers.  In the
extraction of $\Delta E_{nK}$ from the correlation functions, a
correlated $\chi^2$-fit is performed over the fitting interval.
Further, in the extractions of the two- and three- body scattering
parameters, correlations among the different $\Delta E_{nK}$ are
included by performing a coupled $n$- and $t$- correlated fit.
Systematic uncertainties are assigned to these extractions by varying
the ends of the fitting intervals.  It is interesting to note that the
uncertainty in the effective $n$-kaon energy grows with time. This is
expected when $m_\pi+m_\phi<2m_K$ (as is the case for the ensembles
studied here) based on the general arguments given in
Ref.~\cite{Lepage:1989hd}.

%%%%%%%%%%%%%%%%%%%%%%%%%%%%%%%%%%%%%%%%%%%%%%%%%%%
\section{Three-Meson  interactions}
\label{sec:threebody}

\begin{figure}[!ht]
  \centering
  \includegraphics[width=0.45\columnwidth]{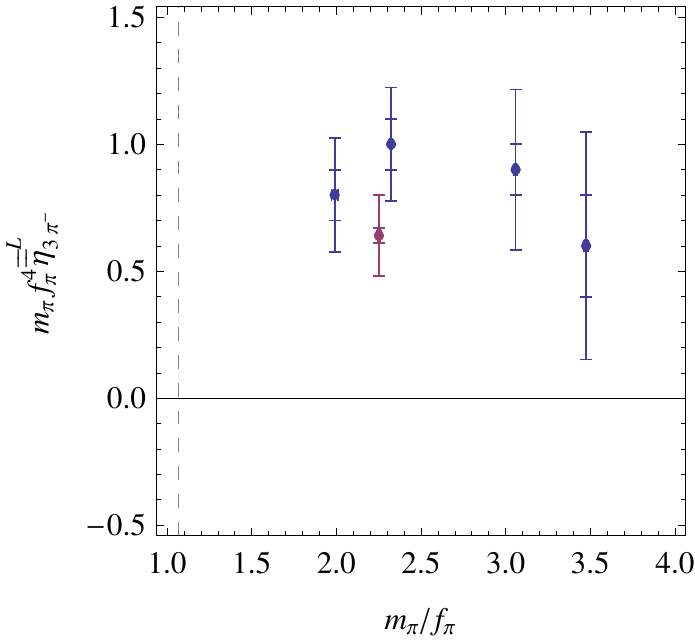}
\qquad
  \includegraphics[width=0.45\columnwidth]{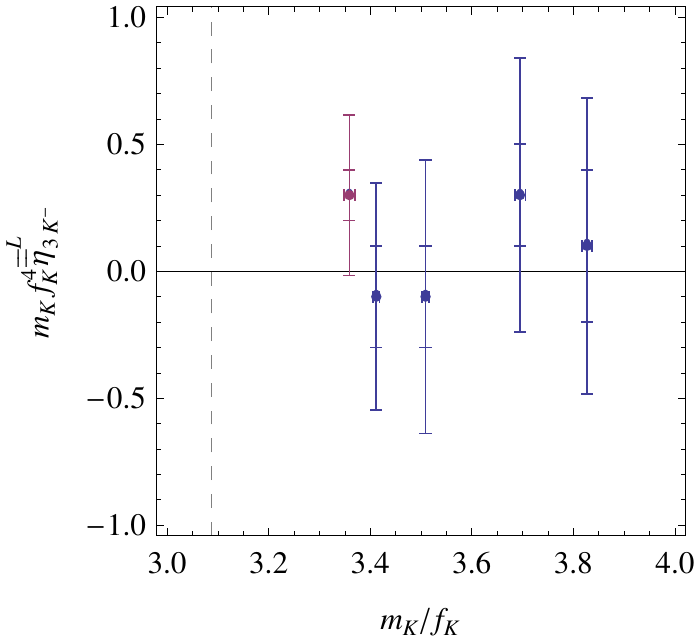}
  \caption{ The three-body interactions,  $m_{\pi} f_{\pi}^4\ 
    \overline{\overline{\eta}}^L_{3\pi^-}$ and $m_{K} f_{K}^4\ 
    \overline{\overline{\eta}}^L_{3K^-}$, defined in
    eq.~(\ref{eq:energyshift}), as a function of $m_\pi/f_\pi$ and
    $m_{K}/f_{K}$, respectively.  The statistical uncertainties and
    the statistical and systematic uncertainties combined in
    quadrature are shown as the inner and outer error-bounds,
    respectively.  The dashed vertical lines indicates the physical
    point.  }
  \label{fig:threebodysummary}
\end{figure}

\noindent
The three-meson interactions extracted from our calculations, $m_{\pi}
f_{\pi}^4\ \overline{\overline{\eta}}^L_{3\pi^-}$ and $m_{K} f_{K}^4\ 
\overline{\overline{\eta}}^L_{3K^-}$, are shown in
Fig.~\ref{fig:threebodysummary}, in units of the NDA estimate,
$\overline{\overline{\eta}}^{L,(NDA)}_{3\pi^-} = 1/(m_\pi f_\pi^4)$
and $\overline{\overline{\eta}}^{L,(NDA)}_{3K^-} = 1/(m_K f_K^4)$.
Table~\ref{tab:results} contains the extracted values of the two-pion
and two-kaon scattering parameters and the three-pion and three-kaon
interactions for each ensemble.
The results of the calculation of the three-pion interaction on the
coarse MILC lattices have been presented previously, and the result
from the fine ensemble is consistent with the closest mass coarse
results, as can be seen in fig.~\ref{fig:threebodysummary}.  The
three-kaon interaction is found to be consistent with zero for all the
kaon masses that have been explored, but the uncertainties are
moderately large.  This result is consistent with the trend found in
multi-pion systems where there is a hint that the three-pion
interaction is decreasing with increasing pion mass.  As is the case
for the three-pion interaction~\cite{Detmold:2008fn}, it is not yet
possible to extract the RGI three-body interaction,
$\overline{\eta}_{3K}^L$ or the underlying parameter $\eta_{3K}(\mu)$
because of the current uncertainties in the calculation.  The
uncertainties of the three-body interactions measured on the large
volume ensemble are significantly larger than those shown in the
figure and we do not display them.

%%%%%%%%%%%%%%%%%%%%%%%%%%%%%%%%%%%%%%%%%%%%%%%%%%%
%
%               TABLE: Lattice Parameteres
%
%%%%%%%%%%%%%%%%%%%%%%%%%%%%%%%%%%%%%%%%%%%%%%%%%%%
\begin{table}[t]
 \caption{Extracted values of the two-body and three-body interactions.
Note that  $\overline{\overline{\eta}}^L_{3\pi^-}$ and 
   $\overline{\overline{\eta}}^L_{3K^-}$ depend on
   $L$, eq.~(\ref{eq:eta3barbar}) and eq.~(\ref{eq:etathreebar}) and are
   quoted at their physical volumes. The
   systematic uncertainties in the pionic quantities from the coarse
   $L\sim 2.5~{\rm fm}$  ensembles have been symmetrized about the central 
value~\protect{\cite{Detmold:2008fn}}. 
The   pion and kaon scattering lengths obtained on the  coarse
   $L\sim 2.5~{\rm fm}$  ensembles are in agreement with those of
   Ref.~\protect{\cite{Beane:2007uh,Beane:2007xs}}, but have  slightly smaller uncertainties
   stemming from increased statistics and from the $n$-correlated
   analysis performed here.
Quantities extracted from the correlators formed from the P$\pm$A propagators are
consistent with those from the individual propagators. 
}
\label{tab:results}
\begin{ruledtabular}
\begin{tabular}{ccccc}
 Ensemble        
&  $m_\pi\overline{a}_{\pi\pi}^{(I=2)}$ &
$m_\pi f_\pi^4\overline{\overline{\eta}}^L_{3\pi^-}$ 
&  $m_K\overline{a}_{KK}^{(I=1)}$ &
$m_K f_K^4 \overline{\overline{\eta}}^L_{3K^-}$ \\ 
\hline 
2064f21b676m007m050 & 0.164(4)(8) & 0.8(1)(2) & 0.491(9)(13) & -0.1(2)(4) \\
2064f21b676m010m050 & 0.206(5)(6) & 1.0(1)(2) & 0.503(11)(19) & -0.1(2)(5) \\
2064f21b679m020m050 & 0.350(7)(10) & 0.9(1)(3) & 0.536(10)(19) & 0.3(2)(5)  \\
2064f21b681m030m050 & 0.476(10)(16) & 0.6(2)(4) & 0.584(12)(18) & 0.1(3)(5) \\
\hline
2896f2b709m0062m031 & 0.178(2)(8) & 0.64(3)(16) & 0.402(6)(18) & 0.3(1)(3)\\
\hline
2864f2b676m010m050 &  0.25(1)(6) & 0(1)(1) & 0.52(6)(17) & -5(3)(8) \\
\end{tabular}
\end{ruledtabular}
\end{table}
%%%%%%%%%%%%%%%%%%%%%%%%%%%%%%%%%%%%%%%%%%%%%%%%%%%

%%%%%%%%%%%%%%%%%%%%%%%%%%%%%%%%%%%%%%%%%%%%%%%%%%%%%%%%
\section{The Equation of State and Chemical
  Potentials}
\label{sec:chemical}

\begin{figure}[!t]
\vskip0.5in
\centering
\includegraphics[width=1.0\textwidth]{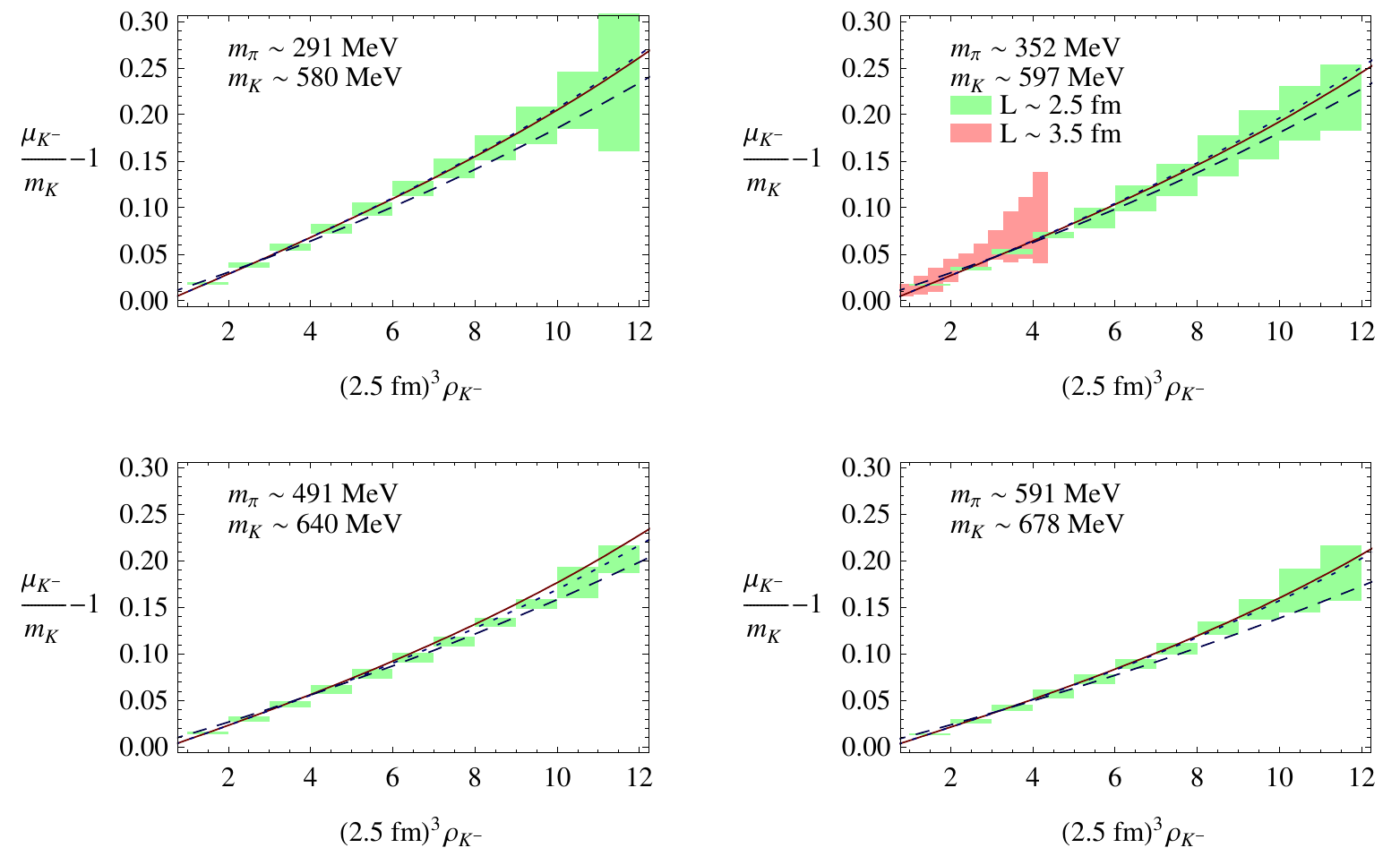} 
\caption{\label{fig:chemicalK} 
  The $K^-$ chemical potential, $\mu_{K^-}$, as a function of the
  density of $K^-$'s on the coarse MILC lattices.  The rectangular
  regions are the results of the lattice calculation where a
  finite-difference has been used to construct the derivative with
  respect to the number of $K^-$'s.  The lighter (green) rectangles
  are obtained from the lattices with spatial extent $L\sim 2.5~{\rm
    fm}$, while the darker (red) rectangles are obtained from the
  lattices with $L\sim 3.5~{\rm fm}$.  The dashed curves correspond to
  tree-level $\chi$PT.  The darker solid curve corresponds to the
  analytic expression for the energy of the ground state in the large
  volume expansion, eq.~(\protect\ref{eq:energyshift}), using the fit
  values for $\aKK$ and \sheep . The lighter dotted curve shows the
  analytic forms with the fit value of $\aKK$ but with \sheep set to
  zero.  }
\end{figure}

\begin{figure}[!t]
\vskip0.5in
\centering
\includegraphics[width=1.0\textwidth]{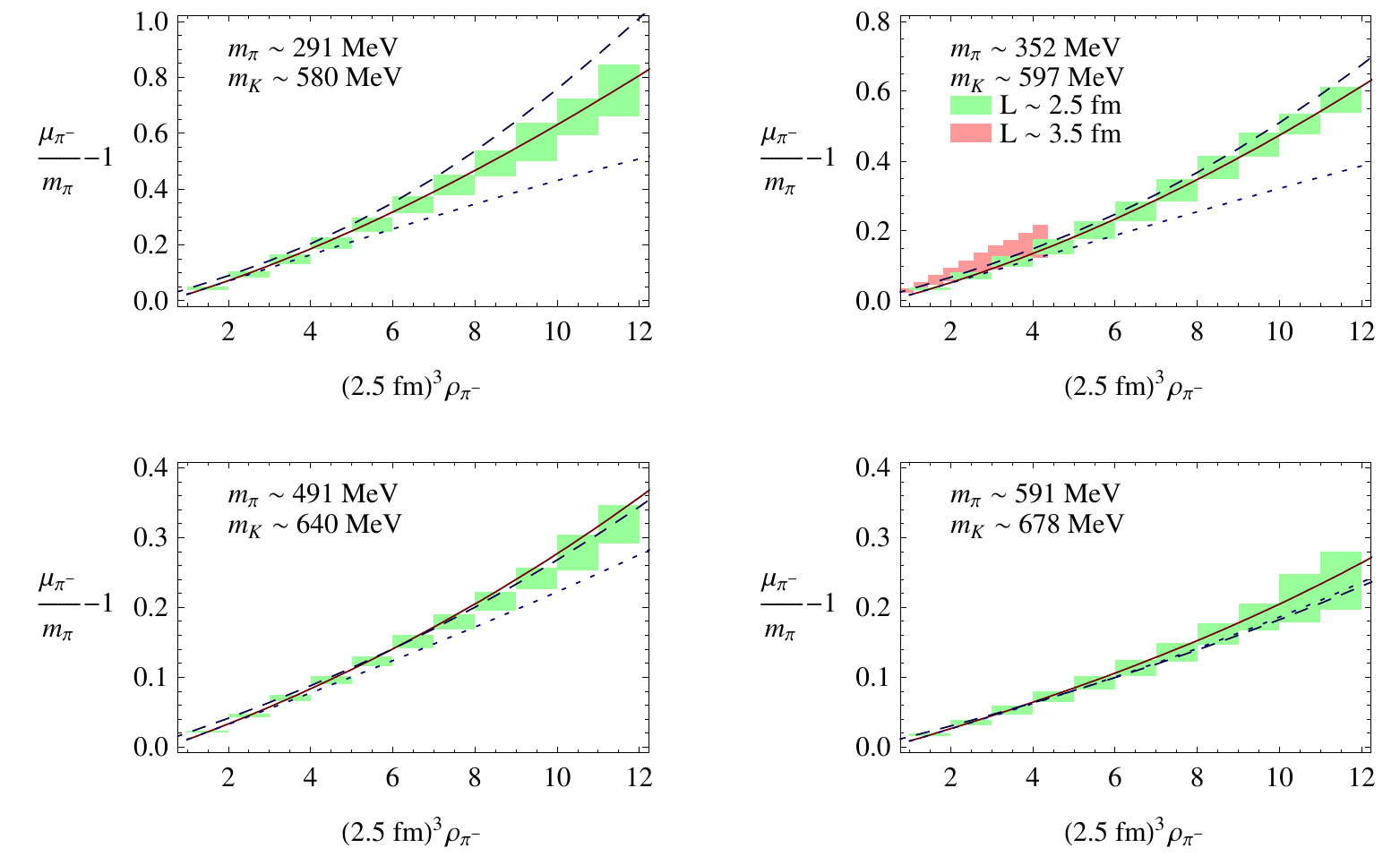} 
\caption{\label{fig:chemicalpi} 
  The isospin-chemical potential, $\mu_{I}$, as a function of the
  density of $\pi^-$'s on the coarse MILC lattices.  The rectangular
  regions are the results of the lattice calculation where a
  finite-difference has been used to construct the derivative with
  respect to the number of $\pi^-$'s.  The lighter (green) rectangles
  are obtained from the lattices with spatial extent $L\sim 2.5~{\rm
    fm}$, while the darker (red) rectangles are obtained from the
  lattices with $L\sim 3.5~{\rm fm}$.  The dashed curves correspond to
  tree-level $\chi$PT.  The darker solid curve corresponds to the
  analytic expression for the energy of the ground state in the large
  volume expansion, eq.~(\protect\ref{eq:energyshift}), using the fit
  values for $\apipi$ and \sheeppi . The lighter dotted curve shows
  the analytic forms with the fit value of $\apipi$ but with \sheeppi
  set to zero.  }
\end{figure}

\begin{figure}[!t]
\vskip0.5in
\centering
\includegraphics[width=0.457\textwidth]{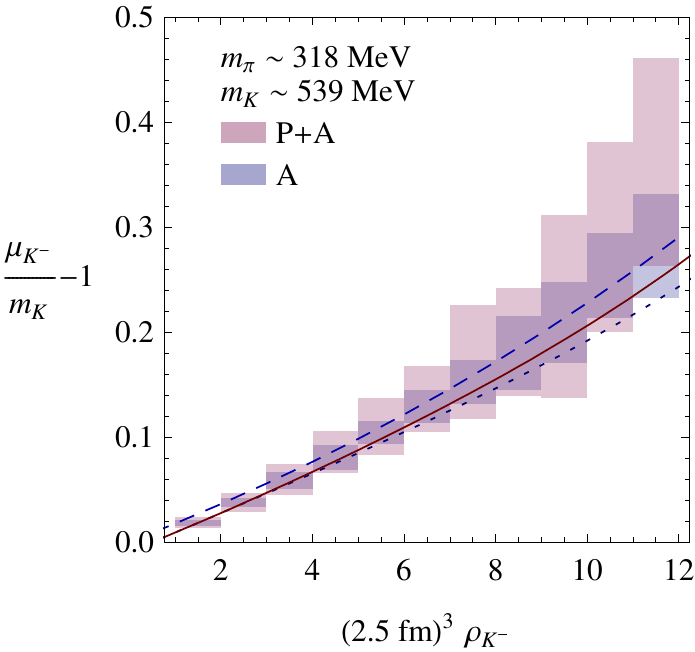}
\qquad
\includegraphics[width=0.45\textwidth]{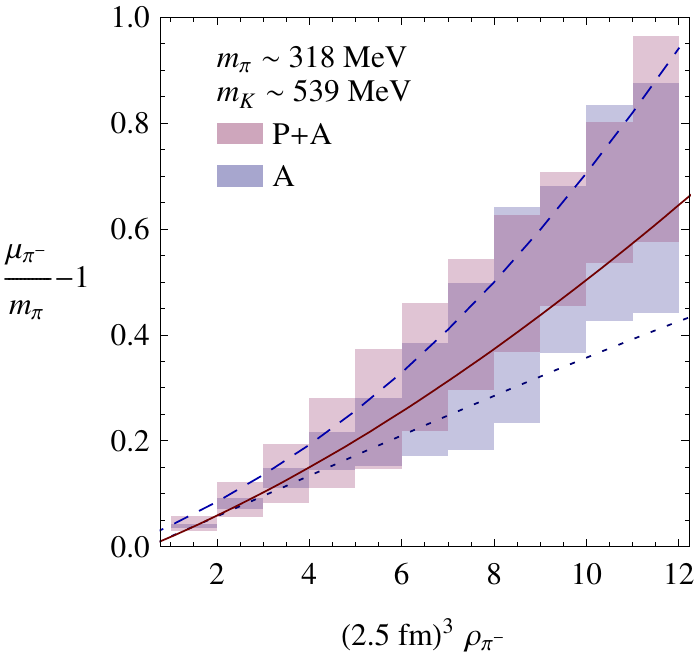} 
\caption{\label{fig:chemicalKfine} 
  The $K^-$ chemical potential, $\mu_{K^-}$, as a function of the
  density of $K^-$'s on the fine MILC lattice ensemble (left panel)
  and the isospin chemical potential, $\mu_I$ as a function of the
  density of $\pi^+$'s on the same ensemble (right panel).  The
  rectangular regions show the results of the lattice calculation
  where a finite-difference has been used to construct the derivative
  with respect to the density of mesons.  The lighter (blue)
  rectangles are obtained from purely anti-periodic temporal boundary
  conditions on the quark propagators, while the darker (purple)
  rectangles are obtained from the P$\pm$A propagators.  The dashed
  curves correspond to LO $\chi$PT.  The darker solid curve
  corresponds to the analytic expressions for the energy of the ground
  state in the large volume expansion,
  eq.~(\protect\ref{eq:energyshift}), using the fit values of the
  two-body and three-body interactions. The lighter dotted curve shows
  the analytic forms with the fit value of the two-body interaction
  but with three-body interactions set to zero.  }
\end{figure}

\noindent
The energy of the $n$ $K^-$ system as a function of volume and of the
number of $K^-$'s is given explicitly in eq.~(\ref{eq:energyshift}) in
the large-volume expansion.  From the equation of state, the $K^-$
chemical potential is
\begin{eqnarray}
  \label{eq:Kplusdensity}
  \mu_{K^-} &=& \left.\frac{d E_{n K}}{d n}\right|_{V={\rm const}} 
  \ \ \ ,
\end{eqnarray}
and can be constructed analytically from eq.~(\ref{eq:energyshift}) or
numerically from the results of the lattice calculations by using a
simple finite difference approximation.  The resulting ratios of the
$K^-$-chemical potentials to the kaon mass on each of the coarse
ensembles are shown in Fig.~\ref{fig:chemicalK} as a function of the
$K^-$-density, $\rho_{K^-}$ in units of (2.5 fm)$^3$.  For comparison
Fig.~\ref{fig:chemicalpi} shows the isospin chemical potential
computed from the $n$ $\pi^+$ systems, expanding upon the results of
Ref.~\cite{Detmold:2008fn} by including the large volume coarse
ensemble.  Finally, Fig.~\ref{fig:chemicalKfine} shows the kaon and
pion chemical potentials for the fine ensemble, and a comparison
between results obtained with anti-periodic boundary conditions and
those computed with the P$\pm$A propagators.  We note that for the
$12$-$K^-$ system, the number density is $\rho_{K^-}^{(12)}=12/L^3 =
0.77~{\rm fm}^{-3}$ on the ensembles with $L\sim 2.5~{\rm fm}$.  The
solid curves in Figs.~\ref{fig:chemicalK}, \ref{fig:chemicalpi}
and~\ref{fig:chemicalKfine} correspond to the prediction at ${\cal
  O}(L^{-7})$ from eq.~(\ref{eq:energyshift}) using the extracted
two-body and three-body scattering parameters, and differ
insignificantly from that at ${\cal O}(L^{-6})$. The dotted curves
result from setting the three-body interactions to zero.

The results of the lattice QCD calculation are consistent within
uncertainties with the prediction of tree-level $SU(3)_L\otimes
SU(3)_R$ $\chi$PT which are shown as the dashed lines in
Figs.~\ref{fig:chemicalK}, \ref{fig:chemicalpi}
and~\ref{fig:chemicalKfine}.  Given that one expects the tree-level
result to be accurate at the $\sim 25\%$-level, this is quite a
remarkable result.  The contributions from the higher-order
counterterms and loop diagrams have conspired to produce a shift from
tree-level that is much less than one expects from NDA.

In comparing these lattice QCD calculations of the chemical potential
to those calculated in $\chi$PT in Section \ref{sec:XPT}, we note that
they are performed in finite volumes and at non-vanishing lattice
spacings.  These approximations modify the chemical potential expected
in the lattice calculation from chiral perturbation theory at most at
NLO in the chiral expansion.  The fact that LO $\chi$PT is in good
agreement with our numerical results, on both the larger and smaller
volume coarse ensembles and the fine ensemble, suggests that finite
volume and lattice spacing effects are small. For $m_\pi\sim352$~MeV,
the two calculations at different volumes agree within their
uncertainties, also supporting this claim.

%%%%%%%%%%%%%%%%%%%%%%%%%%%%%%%%%%%
\section{Conclusions}

\noindent
Kaon condensation may play an important role in the evolution of
supernovae~\cite{Brown:1993jz}.  The theoretical analysis of the
condensation mechanism presently relies upon chiral perturbation
theory to determine both the interactions of kaons with the baryonic
matter and the kaon self-interactions.  We have performed the first
Lattice QCD investigation of the charged kaon condensate, isolating
part of the mechanism that may be in action in the interior of neutron
stars.  In particular, we have explored the $K^-$ chemical potential
as a function of density.  We have also updated our previous analysis
of the charged pion condensate.  Surprisingly, LO $\chi$PT is found to
reproduce the results of the Lattice QCD calculations to within the
uncertainties of the calculation.  This is significantly better than
the $\sim 25\%$-level expected from dimensional analysis.  However,
this is consistent with the findings of other lattice calculations
that higher orders in chiral perturbation theory make contributions
that are much smaller than naively expected, e.g.
Ref.~\cite{NPLQCDreview}, in this range of meson masses.

To move toward exploring the possibility of kaon condensation in a
baryonic background, lattice calculations of systems containing
multiple nucleons and multiple kaons will be required. Only one
dynamical lattice QCD calculation of the two-nucleon system has been
performed to date~\cite{Beane:2006mx}, and exploration of the
multi-nucleon-multi-kaon systems are left for the future.  In addition
to the increased exponential degradation of the signal-to-noise ratio
associated with baryonic correlation functions~\cite{Lepage:1989hd},
``disconnected diagrams'' (in which a quark is created and annihilated
at the source) will contribute to the multi-nucleon-multi-$K^-$
correlation functions, and therefore the computational resources
required to perform such calculations will be considerably larger than
currently available for the study of multi-hadron systems.

%%%%%%%%%%%%%%%%%%%%%%%%%%%%%%%%%%%
\section{Acknowledgments}

\noindent   
We thank Silas Beane, Tom Luu, Assumpta Parre\~no and Aaron Torok for
contributing to the quark propagator calculations required for this
work.  We thank Michael Endres and Steve Sharpe for discussions, and
R.~Edwards and B.~Joo for help with the QDP++/Chroma programming
environment~\cite{Edwards:2004sx} with which the propagator
calculations discussed here were performed.  The computations for this
work were performed at Jefferson Lab, Fermilab, Centro Nacional de
Supercomputaci\'on (Barcelona, Spain), the University of Washington
and the National Energy Research Scientific Computing Center, which is
supported by the Office of Science of the U.S. Department of Energy
under Contract No. DE-AC02-05CH11231. This research was also supported
in part by the National Science Foundation through TeraGrid resources
provided by the National Center for Supercomputing Applications. We
are indebted to the MILC for use of their configurations.  The work of
MJS and WD was supported in part by the U.S.~Dept.~of Energy under
Grant No.~DE-FG03-97ER4014. The work of KO was supported in part by
the U.S.~Dept.~of Energy contract No.~DE-AC05-06OR23177 (JSA) and by
the Jeffress Memorial Trust, grant J-813, DOE OJI grant
DE-FG02-07ER41527 and DOE grant DE-FG02-04ER41302. The work of AW-L
was supported in part by DOE OJI grant DE-FG02-07ER41527 and DOE grant
DE-FG02-93ER40762.

\appendix

%%%%%%%%%%%%%%%%%%%%%%%%%%%%%%%%%%%%%%%%
\section{Effects of temporal boundaries}
\label{sec:effects-temp-bound}

For temporally periodic gauge configurations and (anti-) periodic
quark propagators (or the P$\pm$A propagators), we are in a position
to account for the effects of the temporal boundary condition. As is
well known, a single meson correlation function, generated from
propagators that are subject to periodic or anti-periodic boundary
conditions, does not decay exponentially but is periodic in time and
is given by the sum of forward and backward propagating exponentials,
leading to the behavior
\begin{eqnarray}
  \label{eq:2}
  C_1(t)\stackrel{t\gg1}{\longrightarrow} A_1 \left[e^{-E_1
      t}+e^{-E_1(T-t)}\right] \,,
\end{eqnarray}
where $T$ is the temporal period and the source is defined to be at
$t=0$.  For a two meson correlation function, there are two
contributions; the two mesons can propagate either forward or backward
or one meson can propagate forward and one backward. This leads to a
correlator
\begin{eqnarray}
  \label{eq:3}
  C_2(t)\stackrel{t\gg1}{\longrightarrow} A_{2} \left[e^{-E_2
      t}+e^{-E_2(T-t)}\right] + B_{1,1}\,,
\end{eqnarray}
where the second term is $t$ independent as it is a product of
$\exp[-E_1t]$ and $\exp[-E_1(T-t)]$.  This can further be extended to
the $n$ meson case, where the $n$-meson correlation function is given
by
\begin{eqnarray}
  C_n(t)\stackrel{t\gg1}{\longrightarrow} A_{n} \left[e^{-E_n
      t}+e^{-E_n(T-t)}\right] +\sum_{m=1}^{\lfloor \frac{n}{2}\rfloor} B_{n,m}\left[e^{-E_{n-m}
      t} e^{-E_m(T-t)} + e^{-E_m t} e^{-E_{n-m}(T-t)}\right]\,. 
\nonumber \\
  \label{eq:4}
\end{eqnarray}
We have used the fact that a positive definite transfer matrix
guarantees that the energies occurring in these expressions do not
depend on the correlator they appear in, although the normalizations
$B_{i,j}$ do.  This is only true over physical length scales for the
DW fermion discretization used herein.

In principle, these expressions can be used to extract the energies,
$E_1,\ldots,E_{12}$ from the measured correlations. However, there are
increasing numbers of parameters to determine as $n$ increases (e.g.,
$n=12$ involves eight new parameters) and we have only been able to
successfully perform this analysis up to $n=6$.  The energies
determined with this procedure are consistent with those extracted
from the plateaus in the effective mass plots of the individual
correlation functions, but with somewhat larger uncertainties.

As a further check of our primary analysis, the correlators generated
by combining the propagators subject to periodic boundary conditions
with those subject to anti-periodic boundary conditions, denoted by
P$\pm$A, (whose behaviors are described by eqs.~(\ref{eq:2}),
(\ref{eq:3}) and (\ref{eq:4}) in which $T$ is replaced by $2T$,
significantly reducing the contributions of backward propagating
states) were compared with the corresponding correlation functions
using only anti-periodic boundary conditions for the quark
propagators.  On both the fine and large volume coarse MILC ensembles,
statistical agreement was found for the extracted $n$-$\pi$ and
$n$-$K$ energies for all $n$ studied.  Fig. \ref{fig:chemicalKfine}
compares the two extractions of the $\pi^-$ and $K^-$ chemical
potentials for the fine ensemble.

In the case of a Dirichlet boundary in time, as employed in the
calculation of quark propagators on the coarse $L\sim 2.5~{\rm fm}$
ensembles, the nature of the states reflected from the walls is less
clear and we do not attempt to account for them, instead restricting
our analysis to a region that is well separated from the walls.

%%%%%%%%%%%%%%%%%%%%%%%%%%%%%%%%%%%%%%%%%%%%%%%%%%
%
%               BIBLIOGRAPHY
%
%%%%%%%%%%%%%%%%%%%%%%%%%%%%%%%%%%%%%%%%%%%%%%%%%%


\begin{thebibliography}{99}

%\cite{Brown:1993jz}
\bibitem{Brown:1993jz}
  G.~E.~Brown and H.~Bethe,
%``A Scenario for a large number of low mass black holes in the galaxy,''
  Astrophys.\ J.\  {\bf 423}, 659 (1994).
  %%CITATION = ASJOA,423,659;%%
  
%\cite{Page:2006ud}
\bibitem{Page:2006ud}
  D.~Page and S.~Reddy,
  %``Dense Matter in Compact Stars: Theoretical Developments and Observational
  %Constraints,''
  Ann.\ Rev.\ Nucl.\ Part.\ Sci.\  {\bf 56}, 327 (2006)
  [arXiv:astro-ph/0608360].
  %%CITATION = ARNUA,56,327;%%

%\cite{Beane:2006gf}
\bibitem{Beane:2006gf}
  S.~R.~Beane, P.~F.~Bedaque, T.~C.~Luu, K.~Orginos, E.~Pallante, A.~Parre\~no and M.~J.~Savage
                  [NPLQCD Collaboration],
  %``Hyperon nucleon scattering from fully-dynamical lattice QCD,''
  Nucl.\ Phys.\  A {\bf 794}, 62 (2007)
  [arXiv:hep-lat/0612026].
  %%CITATION = NUPHA,A794,62;%%


\bibitem{KaplanNelson}
  D.~B.~Kaplan and A.~E.~Nelson, preprint HUTP-86/A023;
  %``KAON CONDENSATION IN DENSE NUCLEONIC MATTER,''
  %%CITATION = HUTP-86/A023;%%
%
  %``Strange Goings on in Dense Nucleonic Matter,''
  Phys.\ Lett.\  B {\bf 175} (1986) 57;
  %%CITATION = PHLTA,B175,57;%%
%
  %``Strange Condensate Realignment in Relativistic Heavy Ion Collisions,''
  Phys.\ Lett.\  B {\bf 192}, 193 (1987);
  %%CITATION = PHLTA,B192,193;%%
%
  %``Kaon Condensation In Dense Matter,''
  Nucl.\ Phys.\  A {\bf 479}, 273 (1988);
  %%CITATION = NUPHA,A479,273;%%
%
  %``Kaon Condensation In Heavy Ion Collisions,''
  Nucl.\ Phys.\  A {\bf 479}, 285 (1988);
  %%CITATION = NUPHA,A479,285;%%
%
  %``STRANGE GOINGS ON IN NEUTRON STARS,''
%\href{http://www.slac.stanford.edu/spires/find/hep/www?irn=1967622}{SPIRES entry}
{\it  IN *BERKELEY 1986, PROCEEDINGS, HIGH ENERGY PHYSICS, VOL. 2* 1430-1432. }

%\cite{Beane:2007es}
\bibitem{Beane:2007es}
  S.~R.~Beane, W.~Detmold, T.~C.~Luu, K.~Orginos, M.~J.~Savage and A.~Torok,
  %``Multi-Pion Systems in Lattice QCD and the Three-Pion Interaction,''
  Phys.\ Rev.\ Lett.\  {\bf 100}, 082004 (2008)
  [arXiv:0710.1827 [hep-lat]].
  %%CITATION = PRLTA,100,082004;%%

%\cite{Detmold:2008fn}
\bibitem{Detmold:2008fn}
  W.~Detmold, M.~J.~Savage, A.~Torok, S.~R.~Beane, T.~C.~Luu, K.~Orginos and A.~Parre\~no,
  %``Multi-Pion States in Lattice QCD and the Charged-Pion Condensate,''
  arXiv:0803.2728 [hep-lat] {\it to appear in Phys. Rev. D}.
  %%CITATION = ARXIV:0803.2728;%%


%\cite{Beane:2007qr}
\bibitem{Beane:2007qr}
  S.~R.~Beane, W.~Detmold and M.~J.~Savage,
  %``n-Boson Energies at Finite Volume and Three-Boson Interactions,''
  Phys.\ Rev.\  D {\bf 76}, 074507 (2007)
  [arXiv:0707.1670 [hep-lat]].
  %%CITATION = PHRVA,D76,074507;%%


%\cite{Detmold:2008gh}
\bibitem{Detmold:2008gh}
  W.~Detmold and M.~J.~Savage,
  %``The Energy of n Identical Bosons in a Finite Volume at O(L^{-7}),''
  Phys.\ Rev.\  D {\bf 77}, 057502 (2008)
  [arXiv:0801.0763 [hep-lat]].
  %%CITATION = PHRVA,D77,057502;%%

%\cite{Son:2000xc}
\bibitem{Son:2000xc}
  D.~T.~Son and M.~A.~Stephanov,
  %``QCD at finite isospin density,''
  Phys.\ Rev.\ Lett.\  {\bf 86}, 592 (2001)
  [arXiv:hep-ph/0005225].
  %%CITATION = PRLTA,86,592;%%

%\cite{Son:2000by}
\bibitem{Son:2000by}
  D.~T.~Son and M.~A.~Stephanov,
  %``QCD at finite isospin density: From pion to quark antiquark
  %condensation,''
  Phys.\ Atom.\ Nucl.\  {\bf 64}, 834 (2001)
  [Yad.\ Fiz.\  {\bf 64}, 899 (2001)]
  [arXiv:hep-ph/0011365].
  %%CITATION = YAFIA,64,899;%%

\bibitem{Kogut:2001id}
  J.~B.~Kogut and D.~Toublan,
  %``QCD at small non-zero quark chemical potentials,''
  Phys.\ Rev.\  D {\bf 64}, 034007 (2001)
  [arXiv:hep-ph/0103271].
  %%CITATION = PHRVA,D64,034007;%%


%\cite{Tan:2007bg}
    \bibitem{Tan:2007bg} S.~Tan,
  %``Three-boson problem at low energy and Implications for dilute Bose-Einstein
  %condensates,''
  arXiv:0709.2530 [cond-mat.stat-mech].
  %%CITATION = ARXIV:0709.2530;%%

  
    \bibitem{Bog} N.~N.~Bogoliubov, J.\ Phys.\ (Moscow) {\bf 11}, 23
  (1947).
%%CITATION = JOPYA,11,23;%%

  
    \bibitem{Lee:1957} T.~D.~Lee, K.~Huang, and C.~N.~Yang, Phys.\ 
  Rev.\ {\bf 106}, 1135 (1957)
 
%\cite{Luscher:1986pf}
    \bibitem{Luscher:1986pf} M.~L\"uscher,
%   ``Volume Dependence of the Energy Spectrum in Massive Quantum Field
%   Theories. 
  %2. Scattering States,''
  Commun.\ Math.\ Phys.\ {\bf 105}, 153 (1986).
  %%CITATION = CMPHA,105,153;%%

%\cite{Luscher:1990ux}
    \bibitem{Luscher:1990ux} M.~L\"uscher,
  %``Two particle states on a torus and their relation to the scattering
  %matrix,''
  Nucl.\ Phys.\ B {\bf 354}, 531 (1991).
  %%CITATION = NUPHA,B354,531;%%


%\cite{Beane:2003da}
    \bibitem{Beane:2003da} S.~R.~Beane, P.~F.~Bedaque, A.~Parre\~no
  and M.~J.~Savage,
  %``Two nucleons on a lattice,''
  Phys.\ Lett.\ B {\bf 585}, 106 (2004) [arXiv:hep-lat/0312004].
  %%CITATION = PHLTA,B585,106;%%


%\cite{Renner:2004ck}
\bibitem{Renner:2004ck}
  D.~B.~Renner {\it et al.},
  % ``Hadronic physics with domain-wall valence and improved staggered sea
  % quarks,''
  %
  Nucl.\ Phys.\ Proc.\ Suppl.\  {\bf 140}, 255 (2005).
  %%CITATION = HEP-LAT 0409130;%%

%\cite{Edwards:2005kw}
\bibitem{Edwards:2005kw}
  R.~G.~Edwards {\it et al.},
  % ``Hadron structure with light dynamical quarks,''
  %
  PoS {\bf LAT2005}, 056 (2005).
  %%CITATION = HEP-LAT 0509185;%%


%\cite{Orginos:1999cr}
\bibitem{Orginos:1999cr}
  K.~Orginos, D.~Toussaint and R.~L.~Sugar,
  % ``Variants of fattening and flavor symmetry restoration,''
  %
  Phys.\ Rev.\ D {\bf 60}, 054503 (1999).
  %%CITATION = HEP-LAT 9903032;%%

%\cite{Orginos:1998ue}
\bibitem{Orginos:1998ue}
  K.~Orginos and D.~Toussaint,
  % ``Testing improved actions for dynamical Kogut-Susskind quarks,''
  %
  Phys.\ Rev.\ D {\bf 59}, 014501 (1999).
  %%CITATION = HEP-LAT 9805009;%%


%\cite{Bernard:2001av}
\bibitem{Bernard:2001av}
  C.~W.~Bernard {\it et al.},
  % ``The QCD spectrum with three quark flavors,''
  %
  Phys.\ Rev.\ D {\bf 64}, 054506 (2001).
  %%CITATION = HEP-LAT 0104002;%%


%\cite{Hasenfratz:2001hp}
\bibitem{Hasenfratz:2001hp}
  A.~Hasenfratz and F.~Knechtli,
  % ``Flavor symmetry and the static potential with hypercubic blocking,''
  %
  Phys.\ Rev.\ D {\bf 64}, 034504 (2001).
  %%CITATION = HEP-LAT 0103029;%%
%\cite{DeGrand:2002vu}
\bibitem{DeGrand:2002vu}
  T.~A.~DeGrand, A.~Hasenfratz and T.~G.~Kovacs,
  % ``Improving the chiral properties of lattice fermions,''
  %
  Phys.\ Rev.\ D {\bf 67}, 054501 (2003).
  %%CITATION = HEP-LAT 0211006;%%
%\cite{DeGrand:2003in}
\bibitem{DeGrand:2003in}
  T.~A.~DeGrand,
  % ``Kaon B parameter in quenched QCD,''
  %
  Phys.\ Rev.\ D {\bf 69}, 014504 (2004).
  %%CITATION = HEP-LAT 0309026;%%

%\cite{Durr:2004as}
\bibitem{Durr:2004as}
  S.~D\"urr, C.~Hoelbling and U.~Wenger,
  % ``Staggered eigenvalue mimicry,''
  %
  Phys.\ Rev.\ D {\bf 70}, 094502 (2004).
  %%CITATION = HEP-LAT 0406027;%%



%\cite{Kaplan:1992bt}
\bibitem{Kaplan:1992bt}
  D.~B.~Kaplan,
  %``A Method for simulating chiral fermions on the lattice,''
  Phys.\ Lett.\  B {\bf 288}, 342 (1992)
  [arXiv:hep-lat/9206013].
  %%CITATION = PHLTA,B288,342;%%

%\cite{Shamir:1992im}
\bibitem{Shamir:1992im}
  Y.~Shamir,
  %``The Euclidean spectrum of Kaplan's lattice chiral fermions,''
  Phys.\ Lett.\  B {\bf 305}, 357 (1993)
  [arXiv:hep-lat/9212010].
  %%CITATION = PHLTA,B305,357;%%

%\cite{Shamir:1993zy}
\bibitem{Shamir:1993zy}
  Y.~Shamir,
  %``Chiral fermions from lattice boundaries,''
  Nucl.\ Phys.\  B {\bf 406}, 90 (1993)
  [arXiv:hep-lat/9303005].
  %%CITATION = NUPHA,B406,90;%%

%\cite{Furman:1994ky}
\bibitem{Furman:1994ky}
  V.~Furman and Y.~Shamir,
  %``Axial Symmetries In Lattice QCD With Kaplan Fermions,''
  Nucl.\ Phys.\  B {\bf 439}, 54 (1995)
  [arXiv:hep-lat/9405004].
  %%CITATION = NUPHA,B439,54;%%

%\cite{Shamir:1998ww}
\bibitem{Shamir:1998ww}
  Y.~Shamir,
  %``Reducing chiral symmetry violations in lattice QCD with domain-wall
  %fermions,''
  Phys.\ Rev.\  D {\bf 59}, 054506 (1999)
  [arXiv:hep-lat/9807012].
  %%CITATION = PHRVA,D59,054506;%%

%\cite{Beane:2006mx}
\bibitem{Beane:2006mx}
  S.~R.~Beane, P.~F.~Bedaque, K.~Orginos and M.~J.~Savage,
  %``Nucleon nucleon scattering from fully-dynamical lattice QCD,''
  Phys.\ Rev.\ Lett.\  {\bf 97}, 012001 (2006)
  [arXiv:hep-lat/0602010].
  %%CITATION = PRLTA,97,012001;%%


%\cite{Beane:2006pt}
\bibitem{Beane:2006pt}
  S.~R.~Beane, K.~Orginos and M.~J.~Savage,
  %``The Gell-Mann - Okubo mass relation among baryons from fully-dynamical
  %mixed-action lattice QCD,''
  Phys.\ Lett.\  B {\bf 654}, 20 (2007)
  [arXiv:hep-lat/0604013].
  %%CITATION = PHLTA,B654,20;%%



%\cite{Beane:2006fk}
\bibitem{Beane:2006fk}
  S.~R.~Beane, K.~Orginos and M.~J.~Savage,
  %``Strong-isospin violation in the neutron proton mass difference from
  %fully-dynamical lattice QCD and PQQCD,''
  Nucl.\ Phys.\  B {\bf 768}, 38 (2007)
  [arXiv:hep-lat/0605014].
  %%CITATION = NUPHA,B768,38;%%

%\cite{Beane:2006kx}
\bibitem{Beane:2006kx}
  S.~R.~Beane, P.~F.~Bedaque, K.~Orginos and M.~J.~Savage,
  %``f(K)/f(pi) in full QCD with domain wall valence quarks,''
  Phys.\ Rev.\  D {\bf 75}, 094501 (2007)
  [arXiv:hep-lat/0606023].
  %%CITATION = PHRVA,D75,094501;%%

%\cite{NPLQCDreview}
\bibitem{NPLQCDreview}
S.R. Beane, K. Orginos and M. J. Savage,
 arXiv:0805.4629 [hep-lat].


%\cite{Lepage:1989hd}
\bibitem{Lepage:1989hd}
  G.~P.~Lepage, `The Analysis Of Algorithms For Lattice Field Theory,''
Invited lectures given at TASI'89 Summer School, Boulder, CO, Jun 4-30, 1989.
Published in Boulder ASI 1989:97-120 (QCD161:T45:1989).
  %%CITATION = C89-06-04;%%


%\cite{Beane:2007uh}
\bibitem{Beane:2007uh}
  S.~R.~Beane, T.~C.~Luu, K.~Orginos, A.~Parreno, M.~J.~Savage, A.~Torok and A.~Walker-Loud
                  [NPLQCD Collaboration],
  %``The K+K+ Scattering Length from Lattice QCD,''
  Phys.\ Rev.\  D {\bf 77}, 094507 (2008)
  [arXiv:0709.1169 [hep-lat]].
  %%CITATION = PHRVA,D77,094507;%%

\bibitem{Beane:2007xs}
  S.~R.~Beane, T.~C.~Luu, K.~Orginos, A.~Parreno, M.~J.~Savage, A.~Torok and A.~Walker-Loud,
  %``Precise Determination of the I=2 pipi Scattering Length from Mixed-Action
  %Lattice QCD,''
  Phys.\ Rev.\  D {\bf 77}, 014505 (2008)
  [arXiv:0706.3026 [hep-lat]].
  %%CITATION = PHRVA,D77,014505;%%

%\cite{Edwards:2004sx}
  \bibitem{Edwards:2004sx} R.~G.~Edwards and B.~Joo [SciDAC
Collaboration],
  %``The Chroma software system for lattice QCD,''
Nucl.\ Phys.\ Proc.\ Suppl.\ {\bf 140} (2005) 832
[arXiv:hep-lat/0409003].
  %%CITATION = NUPHZ,140,832;%%



\end{thebibliography}
\end{document}